\documentclass[a4paper,10pt]{article}

\begin{document}

\title{The Accelerated Expansion of the Universe Challenged by an Effect of the Inhomogeneities. A Review}
 
\author{{\bf Marie-No\"elle C\'el\'erier} 
\\
Laboratoire Univers et TH\'eories (LUTH)  \\
Observatoire de Paris-Meudon \\
5 place Jules Janssen, Meudon, 92195 cedex, France \\
E-mail: marie-noelle.celerier@obspm.fr}

\maketitle

\begin{abstract}

Since its decovery during the late 90's, the dimming of distant SN Ia apparent luminosity has been mostly ascribed to the influence of a mysterious dark energy component. Formulated in a Friedmannian cosmological modelling framework based upon the cosmological ``principle'' hypothesis, this interpretation has given rise to the ``concordance'' model. However, a caveat of such a reasoning is that the cosmological ``principle'' derives from a philosophical Copernican assumption and has never been tested. Furthermore, a weakness of its conclusion, i. e., the existence of a negative-pressure fluid or a cosmological constant, is that it would have profound implications for the current theories of physics. This is why we have proposed a more conservative explanation, ascribing the departure of the observed universe from an Einstein-de Sitter model to the influence of inhomogeneities. This idea has been independently developed by other authors and further enlarged to the reproduction of different cosmological data. We review here the main proposals which have been put forward to deal with this purpose and present some prospects for future developments.

\end{abstract}

% \keywords{Cosmology; Large Scale Inhomogeneities; Dark Energy} 

\maketitle

\section{Introduction}\label{sec1}

Since its decovery during the late 90's \cite{R98,P99}, the dimming of distant SN Ia apparent luminosity has been mostly ascribed to the influence of a mysterious dark energy component \cite{PR03} which is supposed to drive an acceleration of the Universe expansion. 

Based upon the cosmological ``principle'' hypothesis and coupled to an analysis of the other available cosmological data, this interpretation has given rise to the ``concordance'' or $\Lambda$CDM model, developed in the context of a Friedmann-Lema\^itre cosmology. However, a drawback to this reasoning is that the cosmological ``principle'' derives from a philosophical Copernican assumption and has still never been tested. In fact, a test has been proposed by C\'el\'erier \cite{MNC00,MNC05}. This test, which was designed to confront the assumption of homogeneity plus cosmological constant up to the higher redshifts attained by the SN Ia observations, might seem actually challenged by the hypothesis of dark energy with a varying with redshift equation of state. But this last hypothesis looks like adding epicycles to an already inadequate solution.

The dark energy component, a cosmological constant or a negative pressure fluid, represents, in the current ``concordance'' model, about 73\% of the energy density of the Universe. However, a cosmological constant is usually interpreted as the vacuum energy of which current particle physics cannot explain such an amplitude and a negative pressure fluid remains a mysterious phenomenum. This is known as the cosmological constant problem.

Another feature of the luminosity distance-redshift relation infered from the supernova data and analysed in the Friedmannian framework is to yield a late-time acceleration of the expansion rate, about the epoch when structure formation enters the nonlinear regime. This would imply that we live at a time when the matter density energy and the dark energy are of the same order of magnitude. This is known as the coincidence problem. 

Since general relativity has only been tested up to scales of order a planetary system, a second type of explanation has been proposed which implies a modification of this theory at larger distance scales  \cite{C04C06}.

Our current purpose is to review the works dedicated to a more simple and natural proposal, which makes only use of known physics and phenomena. Since it appears that the onset of apparent acceleration and the beginning of structure formation in the Universe are concomitant, the idea that the SN Ia observations could be reproduced by the effect of large scale inhomogeneities has been put forward. This interpretation has been first proposed independently by a few authors \cite{MNC00,DH98,PS99,KT00}, shortly after the release of the data. Then, after a period of relative disaffection, it has experienced a reniewed interest.

Another natural explanation of the observed dimming of the SN Ia is that it might be due to the effect of an actual geometrical cosmological constant. Its value such as it appears in the $\Lambda$CDM model has been indeed predicted in the framework of scale relativity, as $\Lambda = 1.36$ $10^{-56} cm^{-2}$, long before the release of the supernova results \cite{LN93,LN96}. However, we shall not consider this approach here since it is beyond the scope of this review.

In Sec. \ref{meth}, the different methods which have been developed and used to deal with both the cosmological constant and the coincidence problems are presented and discussed. The physical quantities studied in this framework are analysed in Sec. \ref{physqu}. The main results are reviewed in Sec. \ref{physres}. Section \ref{cp} is devoted to the conclusion and the discussion of some prospects.

\section{Presentation and discussion of the methods} \label{meth}

It is not obvious that one can keep using Friedmann-Lema\^itre-Robertson-Walker (FLRW) models to interpret the high precision observational data collected in a regime where most of the mass is clumped into or is forming structures. Three main physical effects might actually be missing \cite{BM06}: (i) the overall (average) dynamics of such a universe could be significantly different from the FLRW one. This effect is usually taken into account by calculating so-called backreaction terms added to effective equations for the dynamical evolution of the physical quantities under study, (ii) light propagation in a clumpy universe might be different from that in a homogeneous one and the luminosity distance-redshift relation could hence be affected, (iii) the fact that we have only one observer could influence the results, since we might live in an under or over-dense region which should induce significant corrections in the data interpretation.

What is the most subtle point of this issue is that what is observed in the supernova data is {\it not} an accelerated universe expansion (this is only an artefact due to the a priori assumption that our universe, even in the local region of structure formation, can be represented by a FLRW model) but only a dimming of the SN Ia luminosity taken into account by the infered luminosity distance-redshift relation.

However, many authors have focussed their attention on effect (i) and backreaction has been studied with two types of methods: (a) in the linear regime of small amplitude fluctuations, perturbative expansions which were subsequently averaged out, (b) for the more general problem of the dynamics induced by all scale inhomogeneities including the nonlinear regime, spatial averaging of nonperturbed models. The local dynamics has also been analysed using different exact solutions of the general relativity equations.

Effects (ii) and (iii) have been mostly studied using toy models constructed with such a kind of exact nonaveraged inhomogeneous solutions. 

In the remainder of the present section, we are going to present each of these methods and discuss their application domain and their validity.

\subsection{Spatial averaging} \label{spataver}

Spatial averaging aims at obtaining the impact of a given inhomogeneity profile upon the assumed large scale FLRW background in terms of backreaction terms added to the Friedmannian evolution equations of scalar quantities such as the expansion rate, the energy density, the isotropic pressure. These backreaction terms can be assimilated to a black energy component if they exhibit the right properties. 

In general relativity, spatial averaging is very much involved since the equations which determine the metric tensor and the quantities calculated from it are highly nonlinear. However, when modelling the Universe, the usual method is to use continuous functions representing, e. g., energy-density, pressure, or other kinematical scalars of the velocity field, implicitly assuming that they represent volume averages of the corresponding fine-scale quantities. But we know that our local Universe is highly inhomogeneous at the scales of planetary systems, stars, galaxies, clusters, super-clusters, etc., and there are hints that this inhomogeneity might extend up to currently unknown distances. Anyhow, the scale, i. e., the size of the volume, over which the averagings are performed are never explicitly defined, while the results of this process obviously depend on it.

Moreover, a volume average is a well-defined quantity for scalars only. For vectors, and all the more for tensors, it leads usually to non-covariant quantities. Some authors have tried to propose exact and covariant definitions for tensor averages (see, e. g., Krasi\'nski \cite{AK97} for a review), but the results of these works are still not conclusive and have not been used in the here reviewed  papers.

Another drawback is that a gauge problem arises when relating the ``true'' and the averaged metric. Scalar quantities only are invariant under coordinate transformations, not tensors. 

One must also be aware that, in a generic spacetime, there are no preferred time-slice one could average over and the results depend on the choice of the hypersurfaces on which this average is performed.

But the main issue is the non-commutating property of the two operations: averaging the metric and calculating the Einstein tensor. In other words, the Einstein tensor calculated from an averaged metric and energy-momentum tensor is not equal to the Einstein tensor first calculated from the fine-scale metric and energy-momentum tensor, then averaged. As a consequence, since the Einstein field equations have only been verified experimentally to hold on the scale of planetary systems, they might not hold on larger scales which require averaging. However, in the standard cosmological approach, the Universe is modelled by adopting exactly the wrong method: take a metric which is assumed already averaged, calculate the corresponding Einstein tensor and equate it to an already averaged energy-momentum tensor. 

This problem and the need for correcting the Einstein equations were brought to general attention by Ellis \cite{GE84}, although papers dealing with this issue had been published long before \cite{SF62}. The question raised by the method consisting of determining the parameters of an a priori assumed FLRW model from observational data is the ``fitting problem'' dicussed by Ellis and Stoeger \cite{ES87}. They stress that the conjecture that the standard model is equivalent to an averaged inhomogeneous model cannot be held. The departure of the ``real'' Universe from the averaged one is usually known as the ``backreaction'' effect. The study of backreaction can be implemented by two methods: either one tries to obtain directly the equations satisfied by the averaged quantities, with minimum assumptions as regards an underlying background, or one takes as a background an homogeneous and isotropic FLRW model and analyses the effect of linear perturbations on this background. In this section, we consider the first approach, the second one being studied in Sec. \ref{pa}.

\subsubsection{Kinematical and dynamical backreactions}

The nonperturbative issue has been delt with by Buchert and collaborators for simplified inhomogeneous cosmological models with an irrotational perfect fluid as the gravitational source. First, Buchert and Ehlers \cite{BE97} have proposed a simple averaging procedure which have been used by Palle \cite{DP02} to deal with the cosmological constant problem. 

Then, Buchert \cite{TB00,TB01} has developed another procedure aimed at constructing an ``effective dynamics'' of spatial portions of the Universe from which observable average characteristics can be infered like the Hubble constant, the effective 3-Ricci scalar curvature and the mean density (and isotropic pressure) of a spatial domain bounded by the limits of observation. In the case where the extension of this simply-connected spatial domain to the whole Universe is possible, such a description may allow to draw conclusions about its global properties.

Relations between average scalar sources (energy density, pressure) and an average scalar geometric quantity, the expansion rate, have thus been derived for a well adapted foliation of spacetime. They involve domain dependent backreaction terms that have been splitted into a ``kinematical'' backreaction consisting of the difference between two terms, the variance of the expansion rate and the shear, and a ``dynamical'' backreaction, i. e., pressure forces. These equations show that the averaged shear fluctuations tend to increase the expansion rate as do the averaged energy source terms (provided the energy condition holds), while the averaged expansion fluctuations have an opposite effect and therefore work to a stabilization of structures. The dynamical backreaction can do both. Another property of these equations is to show that, in general, the effective density and pressure are coupled to the evolution of the averaged spatial curvature. They obey a conservation law only in the case where the domain curvature evolves like in a spatially flat, ``small'' FLRW cosmology, i. e., the domain represents on average a small FLRW universe with its own domain-dependent parameters.

Two effective forms of a set of averaged equations have been derived which exhibit the form of those pertaining to spatially 3-Ricci flat Friedmannian cosmologies. In general, the system of equations obtained is not closed. There are three evolution equations for three variables: the average scale factor, the effective energy density and the effective pressure. However only two of them are independent. Therefore, an {\it effective} equation of state (not to be mistaken for the equation of state pertaining to the underlying fine-scale inhomogeneous cosmology) is needed to close the system. This means that different such inhomogeneous spacetimes with the same initial averages but a different {\it effective} equation of state evolve differently even as far as average quantities are concerned. The averaging issue for scalar quantities is therefore condensed into the problem of finding this effective equation of state, including kinematical as well as dynamical backreaction terms which measure the departure of the inhomogeneous underlying universe from a standard FLRW cosmology.

Since his paper is meant to provide a basic architecture for applications, the author gives in his Sec. 4 some useful formulas for barotropic fluids and, especially, for the simplest class of matter models, $p = \gamma \epsilon$, which is relevant for many issues. He also writes the above sets of averaged equations for some peculiar models of cosmological interest. We refer the interested reader to the original article \cite{TB01}.

An analogous study performed by the same author with, as a gravitational source, irrotational dust \cite{TB00}, yields relations between the average scale factor, the average curvature and a ``kinematical'' backreaction term, defined as in \cite{TB01}. For this class of models, there is no ``dynamical'' backreaction since no pressure is involved. The averaged scalar curvature is coupled to the kinematical backreaction representing the influence of fluctuations in the matter fields on the effective, spatially averaged, dynamical properties of a spatial region of the Universe. For this simplified case, a small perturbation drives the dynamical system for the averaged fields into another ``bassin of attraction'' implying drastic changes of the volume-averaged cosmological parameters even if the backreaction term is numerically small.

\subsubsection{Smoothing the geometry: ``bare'' and ``dressed'' cosmological parameters} \label{sg}

But this ``kinematical (and dynamical) backreaction'' does not comprise the whole story. An averaging procedure in relativistic cosmology is not complete unless one also averages the {\it geometrical} inhomogeneities. Since geometrical fields are tensorial variables for which possible strategies of averaging are not straightforward, there is some possible ambiguity in how such an average could be implemented.

Buchert and Carfora \cite{BC02} have suggested an answer to this problem and proposed a Lagrangian smoothing of tensorial variables as opposed to their Eulerian averaging on spatial domains. Curvature fluctuations thus turn out to be crucial and may even outperform the effect of kinematical fluctuations. The authors define effective cosmological parameters that would be assigned to the smoothed cosmological spacetime. These parameters are ``dressed'' after smoothing out the geometrical fluctuations. Relations between the ``dressed'' and ``bare'' parameters are derived. The former provide the framework for interpreting observations with a ``Friedmann bias'', i. e., as if the observer was living in a Friedmannian universe. The latter represent the actual inhomogeneous cosmological model, spatially averaged.

In two subsequent articles \cite{BC03a,BC03b}, the same authors have discussed this relation and identified two effects that quantify the difference between ``bare'' and ``dressed'' parameters: the ``curvature backreaction'' and the ``volume effect''.

The regional curvature backreaction is built from scalar invariants of the intrinsic curvature. It features two positive-definite parts, the scalar curvature amplitude fluctuations and fluctuations in metrical anisotropies. Depending on which part dominates, one obtains an under or overestimate of the actual averaged scalar curvature.

The volume effect is due to the difference between the volume of the smoothed region and the actual volume of the bumpy region.

To summarize the physical content of geometrical averaging we can say that, in the smoothed model, the averaged scalar curvature is dressed both by the volume effect and the curvature backreaction effect. The volume effect is expected when comparing two regions of distinct volumes, but with the same matter content, in a constant curvature space. The backreaction term encodes the deviation of the averaged scalar curvature from a constant curvature model, e. g., a FLRW space section.

However, the interpretation of relativistic cosmological parameters is still far from trivial. One must add the issue of averaging on the observer's light cone in which case the above effects interact with the time evolution of the model. Moreover, the authors stress they did not study the smoothing in a dynamical setting. Last but not least, Apostolopoulos et al. \cite{PA06} have indicated that, for an observer located inside a spherically symmetric inhomogeneous configuration but away from its centre, the local Hubble parameter and the local acceleration parameter are functions of the angle between the radial direction and that of observation. This uncovers another subtelty of inhomogeneous cosmology where the volume increase at a given point during the expansion results from averaging over the various directions and might therefore not be the most appropriate for the characterization of this expansion.

\subsection{Perturbative analysis} \label{pa}

Another way of dealing with the backreaction problem is to use the theory of gravitational perturbations in an expanding universe. The perturbative analysis is a method which is employed when the deviations from homogeneity and isotropy are assumed to be ``small''. It has been first proposed by Lifshitz \cite{EL46,LK63}, who wrote down the equations verified by the corrections to background homogeneous models of expanding universes limited to first order. These equations have been explicitly integrated by Sachs and Wolfe \cite{SW67} to obtain density fluctuations, rotational perturbations and gravitational waves. Since then, the cosmological perturbation theory has been extensively developed and applied to describe the growth of structures in the Universe, to calculate the microwave background fluctuations or the spectrum of gravitational waves, and in many other considerations (see Mukhanov et al. \cite{MFB92} for a review).

In dealing with perturbations one considers two spacetimes, the physical, perturbed spacetime and a fictious background spacetime described by a FLRW model. A one-to-one correspondence between points in the background and points in the physical spacetime carries the coordinates labelling the points in the background over into the physical spacetime and defines a choice of gauge. A change in the correspondence, keeping the background coordinates fixed, is called a {\it gauge transformation}, to be distinguished from a {\it coordinate transformation} which changes the labelling of the points in the background {\it and} in the physical spacetime {\it together}. The perturbation in some quantity is the difference between the value it possesses at a point in the physical spacetime and the value at the {\it corresponding point} in the background spacetime. Since we are here interested in considering the influence of inhomogeneities, the perturbed quantity will be the density.

To study the influence of such fluctuations about an homogeneous and isotropic FLRW background cosmology on the expansion rate of the Universe, one must identify a local physical variable which describes this expansion rate, calculate the backreaction of the cosmological perturbations on this variable to a given order, and {\it then} spatially average the result. Note that it is of the upmost importance to avoid the deficient procedure consisting of calculating an ``observable'' from the spatially averaged metric, which, in general, does not give the same result as calculating the spatial average of the observable \cite{WU98,GB02}. As we stressed when examining the bare averaging procedure, it is also crucial to define the hypersurfaces on which the averaging is performed by a clear physical prescription in order to avoid the possibility of being misled by higher order gauge artifacts.

The cleanest available calculation of the effect of density fluctuations on the averaged expansion rate of a matter-dominated universe up to second order in the metric variables has been performed by Kolb et al. \cite{KM05a} for an application to superhorizon scale perturbations, i. e., in the framework of an inflationary and postinflationary universe, in the adiabatic case. Owing to the perturbation development employed, this method is only consistent for the study of the effect of very large scale fluctuations, with small amplitude, e. g., in the linear regime of structure formation \cite{KM05a,BM05,BD06}. It is also sensitive to the order at which the development is performed. Not only new terms are added when going to higher orders, but the averaging procedure changes for the lower order terms.

Moreover smaller scale fluctuations can exhibit amplitudes that make the series diverge. As an example, it has been demonstrated by Notari \cite{AN05}, using the second-order development of Kolb et al. \cite{KM05a}, that, while at early times the contribution of subhorizon inhomogeneous gravitational fields is perturbatively subdominant, the perturbative series is likely to diverge around the time of structure formation, due to the growth of the perturbations. Therefore, in such cases, one must employ a nonperturbative method. 
 
Note that a new formalism for the study of nonlinear perturbations in cosmology, based on a covariant and fully nonperturbative approach, has recently been proposed \cite{LV06} but it has not yet been applied to our purpose.

\subsection{Use of exact solutions of Einstein equations} \label{exsol}

The method consisting in using exact solutions to modelize the inhomogeneities observed in the Universe is the most straightforward and devoided of theoretical pitfalls. It is adapted to represent either strong or weak inhomogeneities.

For mathematical simplification and also to account for the local quasi-isotropy of the CMB as measured on our wordline, most of the retained models exhibit spatial spherical symmetry.

\subsubsection{Lema\^itre-Tolman Bondi (LTB) models} \label{LTB}

A class of spatially spherically symmetrical solutions of Einstein's equations, with dust (pressureless ideal gas) as the source of gravitational energy, was first proposed by Lema\^itre \cite{GL33}. It was later on discussed by Tolman \cite{RT34} and Bondi \cite{HB47}. It will be here refered to as the LTB model.

The LTB line-element, in comoving coordinates ($r,\theta,\varphi$) and synchronous time $t$, is
\begin{equation}
ds^2=-dt^2 + S^2(r,t)dr^2 + R^2(r,t)(d\theta^2 + \sin^2 \theta 
d\varphi^2) , \label{eq:14}
\end{equation}
in units $c=1$.

Einstein's equations, with $\Lambda =$0 and the stress-energy tensor of dust, imply the following constraints upon the metric coefficients:
\begin{eqnarray}
S^2(r,t) &=& {R^{'2}(r,t)\over {1+2E(r)}} , \label{eq:15}   \\
{1\over 2} \dot{R}^2(r,t) &-& {GM(r)\over R(r,t)}=E(r) , \label{eq:16} \\
4\pi \rho (r,t) &=& {M'(r) \over R'(r,t) R^2(r,t)} , \label{eq:17}
\end{eqnarray}
where a dot denotes differentiation with respect to $t$ and a prime, with respect to $r$, and $\rho (r,t)$ is the energy density of the matter. $E(r)$ and $M(r)$ are arbitrary integration functions of $r$. $E(r)$ can be interpreted as the total energy per unit mass and $M(r)$ as the baryonic mass within the sphere of comoving radial coordinate $r$. 

It is easy to verify that Eq.(\ref{eq:16}) possesses solutions for $R(r,t)$, which differ owing to the sign of the $E(r)$ function and run as follows.
\begin{enumerate}
\item with $E(r)>0$, for all $r$
\begin{eqnarray}
R&=&{GM(r)\over 2E(r)} (\cosh u-1) , \label{eq:18} \\
t-t_0(r)&=&{GM(r)\over [2E(r)]^{3/2}}(\sinh u-u) . \nonumber \\
\nonumber
\end{eqnarray}

\item with $E(r)=0$, for all $r$
\begin{eqnarray}
R(r,t)=\left[ {9GM(r)\over 2}\right]^{1/3} [t-t_0(r)]^{2/3} . 
\label{eq:19}
\\ 
\nonumber
\end{eqnarray}

\item with $E(r)<0$, for all $r$
\begin{eqnarray}
R&=&{GM(r)\over -2E(r)}(1-\cos u) ,  \label{eq:20} \\
t-t_0(r)&=&{GM(r)\over [-2E(r)]^{3/2}} (u-\sin u) . \nonumber \\
\nonumber
\end{eqnarray}
\end{enumerate}
where $t_0(r)$ is another arbitrary function of $r$, usually (but improperly \cite{MNC00}) interpreted, for cosmological use, as a Big-Bang singularity surface.

It seems therefore that the evolution of a LTB universe would depend on three arbitrary functions of $r$: $E(r)$, $M(r)$ and $t_0(r)$. However, the metric and the evolution equations are covariant under any arbitrary coordinate transformation $r = f(r')$. Using this coordinate freedom, one can choose to specify arbitrarily the form of one of these three integration functions. The choice of the two remaining ones is therefore sufficient to characterize completely a given LTB model.

Since a dust model seems appropriate to describe our local matter dominated region of the Universe, and owing to its mathematical simplicity, the LTB solution has been widely retained to deal with the inhomogeneity issue. The matching of the data amounts to find a couple of functions of $r$, arbitrarily chosen, able to fit the observations. However, some authors have claimed that unphysical properties might prevent them to be used as toy models. 

Bolejko \cite{KB05} has studied a set of LTB models with a centered observer, confronted them to the CMB and supernova observations and concluded, on philosophical grounds, that, even if some of them support the hypothesis of the absence of dark energy, a centered observer assumption might be too special to allow these models to be retained as appropriate to deal with the issue. 

Vanderveld et al. \cite{VF06} have shown that many of the proposed LTB models contain a weak singularity at the centre and are therefore unphysical if this is the location retained for the observer. Moreover, examining the problem consisting of guessing a Bang-time function $t_0(r)$ that would yield the measured luminosity distance-redshift relation for zero energy (flat) LTB models, they have found singularities in the differential equations determining $t_0(r)$ which limit the range of redshifts for which these models can reproduce general observations of SN Ia. However, they have identified some singularity-free models exhibiting regions which mimic an accelerating Universe, usually at low redshifts. It must be stressed here that a critical redshift above which the solution is no longer valid is not a drawback from the coincidence point of view. It might on the contrary be considered as a nice property of the model, since what is desired is also a transition from accelerated-like to non accelerated-like expansion at some redshift. As a conclusion, the authors do not exclude the possibility of reproducing the supernova observations in a (flat or non flat) LTB model without dark energy but discuss the task of matching all of the other cosmological data whith such a model.

However, in the literature, the observer has been assumed located either at the centre of the model \cite{GS04,AAG06} or offcentre \cite{JM05}. Therefore, the critics linked to his location do not systematically hold.

Another feature of the LTB solution is the occurence of shell-crossings when $R'=0$ and $M' \neq 0$. In this case, the energy density becomes infinite. To avoid this drawback, some constraints on the model parameters have been identified by Hellaby and Lake \cite{HL85} and must be verified by the retained proposals.

\subsubsection{Stephani models} \label{steph}

Another class of exact models has been used in this framework, Stephani solutions. Stephani models are the most general conformally flat, expanding, perfect fluid spacetimes. They exhibit vanishing shear and rotation, but non-zero acceleration and expansion. Although the general Stephani model possesses no symmetry, the only classes considered to deal with the cosmological constant problem have been those with spherical symmetry.

One feature of the Stephani models that has been the subject of much debate is their matter content. Since in these models the density is homogeneous but the pressure is not, the usual perfect fluid interpretation precludes in general the existence of a barotropic equation of state. Individual fluid elements can behave in a rather exotic manner, e. g., exhibiting negative pressure. For this reason, amongst others, Lorentz-Petzold \cite{LP86} has claimed that Stephani models are not viable descriptions of the Universe. However, Krasi\'nski \cite{AK97} and Barrett and Clarkson \cite{BC00} have argued vigorously that this conclusion is incorrect. Moreover, a negative pressure is also exhibited by dark energy.

\subsubsection{Other kinds of models} \label{okm}

The models proposed in Secs. \ref{LTB} and \ref{steph} have been criticized on the ground of their spherical symmetry which has been thought as non-physical by some authors \cite{KB05,VF06}. However, a spherically symmetric model of universe can be considered as an anisotropic inhomogeneous one averaged over angular scales, which is not physically worse than a uniformly averaged universe such as in the FLRW picture.

Anyhow, some authors have proposed models avoiding spherical symmetry and its contested properties. Moffat \cite{JM06a} has studied a peculiar class of the Szafron family of solutions to Einstein equations. Other exact models have consisted of FLRW patch(es) embedded in a FLRW background with different energy densities \cite{KT00,KT01,KT03,SR06a}.

Depending on the authors, these exact solutions have been used either to barely fit the data as they are obtained by observation or to provide the inhomogenous models over which an averaging procedure ``a la Buchert'' have been implemented \cite{SR06a}.

\section{Studied physical quantities} \label{physqu}

\subsection{The deceleration parameter} \label{decpar}

When reasoning in the framework of a Friedmannian cosmology, the dimming of the supernovae is associated with an acceleration of the Universe expansion. This is why a number of authors have focussed on the issue of either demonstrating or ruling out an effect of the inhomogeneities on the expansion rate.

Some of them have tried to derive \cite{HS05,EF05} or rule out \cite{PA06, JM06b} no-go theorems, i. e., theorems stating that a locally defined expansion can never be accelerating in models where the cosmological fluid satisfies the strong energy condition. However, when spatially averaged such as to reproduce a Friedmannian-like behaviour, a physical quantity associated with the expansion rate behaves quite differently \cite{JM06b,JM06c,KK06}. Therefore, it is very difficult to yield general rules from such theorems.

Others have stressed that the definition of a deceleration parameter in an inhomogeneous model is tricky. Hirata and Seljak \cite{HS05} have proposed four different definitions of such a parameter in an inhomogeneous framework. Apostolopoulos et al. \cite{PA06}, examining the effect of a mass located at the centre of a spherically symmetric configuration on the dynamics of a surrounding dust cosmological fluid, have shown that, for an observer located away from the centre, (i) a central overdensity leads to acceleration along the radial direction and deceleration perpendicular to it, (ii) a central underdensity leads to deceleration along and perpendicular to the radial direction. This demonstrates that, even locally, the effect of inhomogeneities on the dynamics of the Universe is not trivial.
  
Ishibashi and Wald \cite{IW06} have also argued that an averaged quantity representing the scale factor or the deceleration parameter may accelerate without there being any observable consequence.

Another pitfall of this method has been pointed out by Romano \cite{AR06}. He has shown that a peculiar LTB model with positive averaged acceleration can require averaging on scales beyond the event horizon of a central observer. In such cases, the averaging procedure do not preserve the causal structure of spacetime and can lead to the definition of locally unobservable average quantities. This reinforce the statement that a positive averaged acceleration of an underlying inhomogeneous Universe is in general not equivalent to a positive acceleration infered from observations within a Friedmannian scheme.

Another unexpected effect has been put forward by Tomita \cite{KT00,KT01,KT03} who has considered a cosmological model composed of a low-density inner homogeneous region connected at some redshift to an outer homogeneous region of higher-density. Both regions decelerate, but since the void expands faster than the outer region, an apparent acceleration is experienced by the observer located inside the void.

We therefore conclude that the computation of some local quantity (generally the deceleration parameter), eventually subsequently averaged, and behaving the same way as in FLRW models with dark energy can lead to spurious results \cite{GB02,SR04b,BC06} and must therefore be avoided. Actually, what we observe is the dimming of the SN Ia. ``Acceleration'' is a mere consequence of the homogeneity assumption.

\subsection{The Effective Stress Energy Tensor} \label{SET}

Martineau and Brandenberger \cite{MB05} have tried to estimate the effect of a backreaction of Super-Hubble modes, by computing it in terms of the second order Effective Stress Energy Tensor (SET) of cosmological perturbation theory. This has been criticized by Ishibashi and Wald \cite{IW06} who have argued that a large SET implies the contribution of higher order terms in the perturbative scheme and can therefore impair the claimed results.

\subsection{The luminosity distance-redshift relation} \label{ldr}

The luminosity distance-redshift relation is the only direct product of the supernova data, obtained without any a priori cosmological assumption. Of course, it is based on the hypothesis that the SN Ia can be retained as accurate standard candles and this has been thoroughly discussed in the literature \cite{R98,P99,DL00}. However, we shall not discuss this point here since it is beyond the scope of this review.

This relation is therefore the best observable to be fitted and this has been the aim of a number of works \cite{MNC00,DH98,PS99,BM06,PA06,KB05,GS04,AAG06,BC00,IN02,DG06,SJ01,CR06,EM06}.

\section{Main physical results} \label{physres}

\subsection{Superhorizon versus subhorizon fluctuations} \label{supsub}

It has been proposed that energy density perturbations of wave-length larger than the Hubble radius, generated during inflation, might produce upon cosmic parameters effects that could mimic an accelerated expansion of the Universe.

Kolb et al. \cite{KM05a,KM05b} and Barausse et al. \cite{BM05} have calculated to second order in cosmological perturbation theory the corrections to the expansion rate evolution of a matter dominated FLRW universe, in the adiabatic case. They have claimed that the local Hubble constant and the ``deceleration'' parameter can exhibit a large cosmic variance, depending on the physical conditions prevaling during inflation, with the strongest contribution issued from superhorizon metric perturbations. In this case, the ``deceleration'' parameter would have a non-zero probability to be negative.

The isocurvature case has been analysed by Martineau and Brandenberger \cite{MB05}. They obtain an EMT of which the energy density sign changes with time in the matter dominated era, therefore possibly outweighing the energy density of the cosmic fluid. Moreover, the equation of state of their EMT quickly converges towards that of a cosmological constant.

It has however been shown by a number of authors \cite{HS05,EF05,GC05}, among whom some of the proponents themselves \cite{KM05c,KM05d}, that actually superhorizon fluctuations cannot produce an accelerated expansion.

Even more, Buchert \cite{TB05}, R\"as\"anen \cite{SR06b} and Kolb et al. \cite{KM05d} have subsequently demonstrated that late-time acceleration of the expansion rate without dark energy must involve perturbations with subhorizon wavelength to be possibly obtained.

\subsection{Overall dynamics} \label{ovdyn}

Working in the framework of an ``Onion'' model which is a LTB solution where the structures are shells of different density, Biswas et al. \cite{BM06} have computed a {\it minimal} overall (average) effect which amounts to a correction in apparent magnitudes at all redshifts of order $\Delta m \simeq 0.15$. The authors discuss different possibilities that could further enhance this effect and mimic dark energy.

\subsection{Averaging corrections} \label{avcor}

This is to be added to the estimation made by Buchert and Carfora \cite{BC03b} of the volume effect alone (the Ricci curvature backreaction has not been considered) in a naive swiss-cheese model obtained by glueing Riemannian balls in place of Euclidean ones in the spatial slice. Its magnitude is found to be 64\%, reducing the necessary dark energy in the concordance model roughly from 70\% to 50\% \cite{TB06}.

However, the above estimates of overall and average effects do not encompass the whole influence of such procedures. We have seen indeed in Sec. \ref{spataver} that other backreactions (kinematical, dynamical, curvature backreactions) can also proceed from averaging and smoothing. 

\subsection{Spatial curvature} \label{spcur}

Buchert \cite{TB05,TB06} has studied the constraints which apply to a dust model of universe if one wants to explain dark energy by the backreaction effect of inhomogeneities. He thus argues that the model must be in a far-from equilibrium state, characterized by strong averaged expansion fluctuations on the global scale and that a large value of the kinematical backreaction entails a substantial (negative) average Ricci curvature on the considered domain. This has been confirmed through a different method by the same author and his collaborators \cite{BL06}.

R\"as\"anen \cite{SR06b}, working in the same framework, has also found that a negative average spatial curvature is mandatory to compensate a backreaction increasing the expansion rate of the model. However, he stresses that, owing to the strong Integrated Sachs-Wolfe effect produced by a spatial curvature, a quick transition from a standard Friedmannian behaviour to an accelerated expansion is desired. Such a transition might be more easily obtained with a backreaction issued from small wave-length modes associated to structure formation.

Paranjape and Sing \cite{PS06} have applied Buchert's equations \cite{TB00} to the averaging of LTB models. They find that the average behaviour of spatial slices is that of acceleration only in the unbound, curvature dominated, class. A similar result has been obtained by Moffat \cite{JM06a,JM06c}.

Even if, as discussed above, the link between the accelerated expansion of an average model and the ability of reproducing the supernova data with the bare underlying inhomogeneous universe is not straightforward, this might provide some clue to the solution of the problem: why do the spatially flat models studied in the literature encounter pathologies or mismatches \cite{VF06}?

Moreover, this result is actually consistent with the fact that the exact inhomogeneous models which seem to reproduce best the observed luminosity distance-redshift relation are those which exhibit negative spatial curvature near the observer, or, equivalently from an average behaviour point of view, where the observer is located in an underdense region \cite{BM06,AAG06,KT01,DG06,SJ01,RM05}. However, Iguchi et al. \cite{IN02} have also found pure Big-Bang (spatially flat) inhomogeneous LTB models which are consistent with the observed relation.

Anyway, the assumption that we might be located within an underdense region seems to be consistent with observations leading to the identification of a Local Void and of its suggested expansion \cite{KT01,KT03,IO04}.

\subsection{Matching cosmological data with exact solutions} \label{matcd}

The results reviewed in the first part of this Sec. \ref{physres} have been obtained in the framework of the backreaction theory which is not already complete. We shall therefore in the following analyse the proposals aiming at reproduce the observations with exact solutions.

The luminosity distance-redshift relation is the observational quantity which much be matched by a cosmological model dedicated to solve the so-called cosmological constant problem. This matching has been successfully performed by a number of proposed models: homogeneous void models \cite{KT01}, LTB models with centered observer \cite{AAG06,IN02,DG06}, LTB models with outcentre observer \cite{BM06,RM05}, Stephani models \cite{DH98,GS04,BC00,SJ01}. Most of the studied models reproduce the $\Lambda$CDM luminosity distance-redshift relation up to a redshift of the order unity. This feature, which has been considered as a ruling out drawback by some authors \cite{KB05,VF06}, must be viewed as a nice way out of the coincidence problem. The physical explanation is that the appearance of small scale inhomogeneities corresponds to the onset of structure formation at the time (around $z \simeq 0.5$) where ``acceleration'' begins to be observed in the FLRW representation of our Universe. Note that such an explanation have also been put forward to justify models of structure formation proposed to solve both the problems of the cosmological constant and of coincidence and analysed with the tools of Buchert's averaging procedure \cite{SR06a}.

Pascual-S\'anchez \cite{PS99} has considered Local Rotational Symmetric inhomogeneous spacetimes, with a barotropic perfect fluid equation of state $p = p(\rho)$ for the cosmic matter which is assumed to verify the null energy condition $\rho +p > 0$. Adding the hypothesis of spatial spherical symmetry of the model, to account for the local isotropy of the cosmic microwave background (CMB) as measured on our wordline, he shows that one can obtain a negative deceleration parameter, covariantly defined \cite{PM84,HM97}, with no need for a cosmological constant or any kind of dark energy. The proceeding cosmic acceleration is due to the presence, in the equations, of a positive inhomogeneity parameter related to the kinematic acceleration or, equivalently, to a negative radial pressure gradient of the cosmic barotropic fluid. 

C\'el\'erier \cite{MNC00} has discussed the possibility of testing the cosmological constant and homogeneity hypotheses up to $z \simeq 1$ by analyzing the coefficients of a Taylor expansion of the observed luminosity distance in powers of the redshift. Since, in the Friemannian picture with cosmological constant, these coefficients are functions of the three cosmological parameters $H_0, \Omega_M$ and $\Omega_\Lambda$, an expansion to at least the fourth order can be used to verify the coherence of the values obtained at the third order. In this paper, the author considers only the case of a cosmological constant. But the generalization to any kind of dark energy with a given equation of state is straightforward. Take an expansion of the equation of state in powers of the redshift, substitute in the Taylor series giving the observed luminosity distance and check the consistency of the values of the coefficients at the necessary order. In the second part of the article, the author proposes, as a mere example of inhomogeneous models able to fit the data, the LTB class with no cosmological constant. She shows that these models are completely degenerate with respect to any magnitude-redshift relation. They can therefore easily reproduce the supernova data, which implies that the apparent acceleration of the local expansion rate might rather be reproduced by many inhomogeneity profiles ascribed to universe models.

An application of these LTB toy models to the fitting of generic $\Lambda$CDM universes has been tried by Iguchi, Nakamura and Nakao \cite{IN02}. The purpose of these authors has been to select peculiar classes of the LTB type which might reproduce the luminosity distance-redshift relation of a FLRW universe with $\Omega_M = 0.3$ and $\Omega_\Lambda = 0.7$. For simplicity, they limit their study to the pure Bang-time and pure curvature inhomogeneity cases. They find that some among these models reproduce consistently the $\Lambda$CDM relations up to $z\simeq 1$, but are unsuccessful to fit it for larger $z$. Even if the more general LTB case was left aside, these results might imply an insight into a possible solution to the coincidence problem. The same kind of study has been performed by Garfinkle \cite{DG06} for a subclass of the LTB models analysed by \cite{IN02}.

A series of papers has been devoted to try to reproduce the cosmological data from the SN Ia with some special spherically symmetric models of the Stephani type where the observer is assumed located at the centre. Initiated by Dabrowski and Hendry \cite{DH98}, a good fit to the data is obtained with a peculiar class of the analysed models and for small redshifts $z \ll 1$ since the magnitude-redshift relation is expanded in powers of $z$ to the first order. In \cite{SJ01}, another class is singled out. It is shown that the high redshift SN Ia fit quite well its $m(z)$ relation without cosmological constant and that the quantity corresponding to a deceleration parameter decreases with the distance becoming very negative for sufficiently far away galaxies. In \cite{GS04}, the magnitude-redshift relation as derived in  \cite{SJ01} is fitted to the SN Ia observations to obtain quantitative estimations of the model parameters. It is also shown that the best fit model is consistent with the location of the three first peaks in the CMB spectrum. Moreover, \cite{DH98} and \cite{SJ01} stress that the age of the Universe for the considered Stephani models is longer than in the Friedmannian counterparts corresponding to the same $H_0$ and $\Omega_0$ parameters.

Barrett and Clarkson \cite{BC00} have also examined the constraints put on the parameters of a special class of Stephani models by some of the  cosmological data for all observer positions. To this end, they have derived exact, analytical expressions for the luminosity distance-redshift relations and the anisotropies issued from the CMB. Even if their main goal has been to challenge the cosmological principle, their results, presented as exclusion plots in the parameter space of the models, show that the SN Ia data can be reproduced in such universes without dark energy.

Mansouri \cite{RM05} has proposed the so-called Structured FRW universe, i. e., a FLRW background with local inhomogeneous patches, grown out of the primordial fluctuations and distributed at random in the background. Each subhorizon local patch is approximated as an inhomogeneous cosmic fluid described by a LTB flat metric. The Bang-time function of the LTB bubbles is interpreted as the time of nucleation of mass condensation in the patch and its behaviour is fixed through the junction conditions at the matter-radiation domination transition. The observer can be located on or off-centre in one of the patches. To avoid a singularity at the centre of the model and shell-crossing, the observer must be located in an underdense region of the patch. The analysis of the luminosity distance-redshift relation shows that, in this toy model, the dimming of distant objects is obtained without any need for dark energy.

Apostolopoulos et al. \cite{PA06} have studied the behaviour of the luminosity function for an observer located at the centre of an overdensity. They assume a continuous distribution of matter, homogeneous in the central region, which falls off at larger distance and asymptotically becomes homogeneous again with an energy density smaller than in the central one. For photons emanating from immediately outside the central overdense region, they observe a strong apparent acceleration, which remains positive up to a certain redshift, then turns negative and asymptotically approachs the value -0.5. For small redshifts, $z \simeq 0$, strong apparent deceleration is found (the authors expect this feature to be modified in more realistic collapse models with pressure). They claim that their results demonstrate the possible link between the growth of perturbations and the perceived acceleration of the expansion. These authors have also studied the model proposed by Tomita \cite{KT00,KT01,KT03}. They have verified that the luminosity distance-redshift relation in this case could be similar to the one in an accelerating homogeneous universe. 

Biswas, Mansouri and Notari \cite{BM06} have proposed an exact simplified model of nonlinear structure formation, the Onion model, which is a LTB solution where the structures are shells of different densities. At early times, around last scattering, the density fluctuations are assumed to have the amplitude observed in the CMB, i. e., the Universe is nearly Einstein-de Sitter (EdS). At redshifts $z \geq {\cal O}(10)$ the density contrast grows exactly as in a perturbed FLRW model. When this density contrast becomes of ${\cal O}(1)$, nonlinear clustering appears. The observer sits in a generic, offcentre, position and looks at sources in the radial direction. An exact expression for the luminosity distance of an object as seen by this observer is derived. The authors show that, even if this model does not yield a significant overall effect (i. e., quantities such as the matter density, on an average, still behave as in the homogeneous EdS universe), corrections to both the redshift and the luminosity distance are significant once the inhomogeneities become large and the nearer the sources get to the observer. Moreover, to explain the mismatch between the measurement of the local Hubble parameter and the luminosity distance of high redshift supernovae, the observer must be located in an underdense region. The authors further show that the Onion model can be consistent with other observations such as local density measurements, the CMB first acoustic peak position (which measures the curvature of the Universe) and the baryon acoustic oscillations.

Alnes, Amarzguioui and Gron \cite{AAG06} have tried to reproduce: (i) at very low redshift, $z<0.12$, the matter density as measured by the 2dF team, $\Omega_{m0} = 0.24 \pm 0.05$; (ii) at low redshift, $0<z<1.5$, the observed dimming of the SN Ia; (iii) at very large $z$, the CMB power spectrum. The models they have retained for this purpose exhibit an underdense region, represented by a LTB solution with negative curvature, centered on the observer and surrounded by a flat, matter dominated universe with no cosmological constant or quintessence, i. e., Einstein-de Sitter. The successes of their best fit model are to reproduce better than with $\Lambda$CDM the supernova data and, approximately, the low $z$ matter density ($\Omega_{m0} = 0.20$ at the centre of the LTB underdensity). However, to obtain the location of the first peak of the CMB power spectrum, they need to assume the background universe to be flat with a value of 0.51 for the Hubble parameter outside the inhomogeneity. An improvement to this model might be to switch the observer off the centre. This is the purpose of some work in progress \cite{AA06}.

Chung and Romano \cite{CR06} have also used LTB models with no cosmological constant to fit the supernova data. They derive an inversion method which allows to obtain the $M(r)$ function for an observed luminosity distance $D_L(z)$, while arbitrarily guessing the form of $E(r)$ and $R(t_0,r)$. Although this method is unstable for redshifts above some cutoff $z_c$, of which the value depends on the retained models, it allows to exactly fit the observations up to this cutoff.

Enqvist and Mattsson \cite{EM06} have analysed some other kinds of LTB solutions. For their best fit model, neither local nor average positive acceleration is needed to match the observations. This reinforce our statement that acceleration is not a mandatory product of the data when they are analysed without an a priori homogeneity assumption. 

Now, it must be stressed once again that the problem of fitting a given luminosity distance-redshift relation with a peculiar inhomogeneous (LTB) model is completly degenerate \cite{MNC00,MH97}. Therefore, the main challenge remains to fit the whole set of available cosmological data. Since very few exact solutions to Einstein's equations can be of use in a cosmological framework, mostly toy models with spherical symmetry have been used up to now to deal with this issue. However, a project to begin implementing this is currently underway \cite{LH06}.

\section{Conclusion and prospects} \label{cp}

One can find in the literature inhomogeneous cosmological toy models able to solve both the cosmological constant and coincidence problems, i. e., to mimic an ``accelerated expansion'' with no need for dark energy up to the epoch when structure formation enters the nonlinear regime, around $z \simeq 1$.

It has been shown that inhomogeneities likely to solve these problems must be of the subhorizon and strong type, which cannot be studied with perturbation methods. Averaging and smoothing procedures have been proposed which can provide some insight into the issue for very simple cases but which much be used with care since they are not devoided of pitfalls and incompleteness.

Exact solutions of Einstein's equations have the nice property of being able to modelize both strong and weak inhomogeneities. What these models must reproduce is {\it not an accelerated expansion}, which is an artefact of the homogeneous assumption, but the observed dimming of the SN Ia luminosity, i. e., the luminosity distance-redshift relation. Some classes have been nicely fitted to this relation at low redshifts. However, the proposed toy models cannot pretend to be fair representations of our neighbouring patch of universe. Some of them exhibit controversed properties issued from their spherical symmetry (LTB models) or their matter content (Stephani models). But since very few exact solutions to Einstein's equations can be of use in a cosmological framework, only simple toy models have been selected up to now to deal with this issue.

Even if some have been shown to match other cosmological observational constraints, the main challenge remains to fit the whole set of available cosmological data. However, a project to begin implementing this issue is currently underway \cite{LH06}. A solution, if any, would be to use non pathological exact inhomogeneous solutions, reproducing the nearby Universe, coupled to or asymptotically matching nearly homogeneous ones valid up to last scattering. 

However, we know that, since we have recently entered the era of precision cosmology, the number and the resolution of the data is going to inflate in the years to come, and we can expect to be able then to reproduce our neighbouring Universe up to the finest scale. Inhomogeneous models will therefore be unavoilable but the challenge will be to choose the ones which will have the required properties to best deal with a given problem.

\section{Acknowledgements} I wish to thank Brandon Carter and Julien Larena for a usefull discussion about the averaging procedures presented in this review.


\begin{thebibliography}{9}

\bibitem{R98} A. G. Riess, A. V. Filippenko, P. Challis, {\it et al},  AJ {\bf 116}, 1009 (1998).

\bibitem{P99} S. Perlmutter, G. Aldering, G. Goldhaber, {\it et al}, ApJ {\bf 517}, 565 (1999).

\bibitem{PR03} see, e.g., P. J. E. Peebles \& B. Ratra, Rev. Mod. Phys. {\bf 75}, 559 (2003).

\bibitem{MNC00} M. N. C\'el\'erier, A \& A {\bf 353}, 63 (2000).

\bibitem{MNC05} M. N. C\'el\'erier, in {\it Proceedings of the 22nd Texas Symposium on Relativistic Astrophysics}, Stanford University, 2004, edited by P. Chen, E. Bloom, G. Madejski \& V. Patrosian, eConf C041213, 1403 (2005).

\bibitem{C04C06} see, e.g., S. M. Carroll, V. Duvvuri, M. Trodden \& M. Turner, Phys. Rev. {\bf D 70}, 043528 (2004); S. Capozziello, {\it ``Dark Energy and Dark Matter as Curvature Effects''}, to be published in the {\it Proceedings of the 11th Marcel Grossmann Meeting}, Berlin, July 2006 (2006).

\bibitem{DH98} M. P. Dabrowski \& M. A. Hendry, ApJ {\bf 498}, 67 (1998).

\bibitem{PS99} J. F. Pascual-S\'anchez, Mod. Phys. Lett. {\bf A 14}, 1539 (1999).

\bibitem{KT00} K. Tomita, ApJ {\bf 529}, 38 (2000).

\bibitem{LN93} L. Nottale, {\it Fractal Space-Time and Microphysics: Towards a Theory of Scale Relativity}, World Scientific, Singapore (1993).

\bibitem{LN96} L. Nottale, Chaos, Solitons \& Fractals {\bf 7}, 877 (1996).

\bibitem{BM06} T. Biswas, R. Mansouri \& A. Notari, arXiv: astro-ph/0606703.

\bibitem{AK97} A. Krasi\'nski, {\it Inhomogeneous Cosmological Models}, Cambridge University Press (1997).

\bibitem{GE84} G. F. R. Ellis, in {\it General Relativity and Gravitation}, edited by B. Bertotii, F. de Felice, \& A. Pascolini, D. Reidel, Dordrecht (1984) p.215.

\bibitem{SF62} M. F. Shirokov \& I. Z. Fisher, Astron. Z. {\bf 39}, 988 (1962) [Sov. Astr. A. J. {\bf 6}, 699 (1963)].

\bibitem{ES87} G. F. R. Ellis \& W. Stoeger, Class. Quant. Grav. {\bf 4}, 1697 (1987).

\bibitem{BE97} T. Buchert \& J. Ehlers, A \& A {\bf 320}, 1 (1997).

\bibitem{DP02} D. Palle, Nuovo Cim. {\bf 117 B}, 687 (2002).

\bibitem{TB00} T. Buchert, Gen. Rel. Grav. {\bf 32}, 105 (2000).

\bibitem{TB01} T. Buchert, Gen. Rel. Grav. {\bf 33}, 1381 (2001).

\bibitem{BC02} T. Buchert \& M. Carfora, Class. Quant. Grav. {\bf 19}, 6109 (2002).

\bibitem{BC03a} T. Buchert \& M. Carfora, Phys. Rev. Lett. {\bf 90}, 031101 (2003).

\bibitem{BC03b} T. Buchert \& M. Carfora, in {\it Proceedings of the 12th JGRG Conference}, Tokyo, 2002, edited by M. Shibata et al. (2003) p. 157.

\bibitem{PA06} P. S. Apostolopoulos, N. Brouzakis, N. Tetradis \& E. Tzavara, JCAP {\bf 06}, 009 (2006).

\bibitem{EL46} E. M. Lifshitz, J. Phys. (Moscow) {\bf 10}, 116 (1946).

\bibitem{LK63} E. M. Lifshitz \& I. M. Khlatnikov, Adv. Phys. {\bf 12}, 185 (1963).

\bibitem{SW67} R. K. Sachs \& A. M. Wolfe, ApJ {\bf 147}, 73 (1967).

\bibitem{MFB92} V. F. Mukhanov, H. A. Feldman \& R. H. Brandenberger, Phys. Rep. {\bf 215}, 203 (1992).

\bibitem{WU98} W. Unruh, ``Cosmological long wavelength perturbations'', arXiv: astro-ph/9802323 (1998).

\bibitem{GB02} G. Geshnizjani \& R. Brandenberger, Phys. Rev. {\bf D 66}, 123507 (2002).

\bibitem{KM05a} E. W. Kolb, S. Matarrese, A. Notari \& A. Riotto, Phys. Rev. {\bf D 71}, 023524 (2005).

\bibitem{BM05} E. Barausse, S. Matarrese \& A. Riotto, Phys. Rev. {\bf D 71}, 063537 (2005).

\bibitem{BD06} C. Bonvin, R. Durrer \& M. A. Gasparini, Phys. Rev. {\bf D 73}, 023523 (2006).

\bibitem{AN05} A. Notari, Mod. Phys. Lett. {\bf A21}, 2997 (2006).

\bibitem{LV06} D. Langlois \& F. Vernizzi, Phys. Rev. Lett. {\bf 95}, 091303 (2005); Phys. Rev. {\bf D 72}, 103501 (2005); JCAP {\bf 02}, 014 (2006).

\bibitem{GL33} G. Lema\^itre, Ann. Soc. Sci. Bruxelles {\bf A 53}, 51 (1933).

\bibitem{RT34} R. C. Tolman, Proc. Nat. Acad. Sci. {\bf 20}, 169 (1934).

\bibitem{HB47} H. Bondi,  MNRAS {\bf 107}, 410 (1947).

\bibitem{KB05} K. Bolejko, arXiv: astro-ph/0512103 (2005).

\bibitem{VF06} R. A. Vanderveld, E. E. Flanagan \& I. Wasserman, Phys. Rev. {\bf D 74}, 023506 (2006).

\bibitem{GS04} W. Godlowski, J. Stelmach \& M. Szydlowski, Class. Quant. Grav. {\bf 21}, 3953 (2004).

\bibitem{AAG06} H. Alnes, M. Amarzguioui \& O. Gron, Phys. Rev. {\bf D 73}, 083519 (2006).

\bibitem{JM05} J. W. Moffat, J. Cosmol. Astropart. Phys. {\bf 0510}, 012 (2005).

\bibitem{HL85} C. Hellaby \& K. Lake, ApJ {\bf 290}, 381 (1985).

\bibitem{LP86} D. Lorenz-Petzold, Astrophys. Astron. {\bf 7}, 155 (1986).

\bibitem{BC00} R. K. Barrett \& C. A. Clarkson, Class. Quant. Grav. {\bf 17}, 5047 (2000).

\bibitem{JM06a} J. W. Moffat, arXiv: astro-ph/0606124 (2006).

\bibitem{KT01} K. Tomita, MNRAS {\bf 326}, 287 (2001).

\bibitem{KT03} K. Tomita, ApJ {\bf 584}, 540 (2003).

\bibitem{SR06a} S. R\"as\"anen, Int. J. Mod. Phys. {\bf D15}, 2141 (2006).

\bibitem{HS05} C. M. Hirata \& U. Seljak, Phys. Rev. {\bf D 72} 083501 (2005).

\bibitem{EF05} E. E. Flanagan, Phys. Rev. {\bf D 71}, 103521 (2005).

\bibitem{JM06b} J. W. Moffat, JCAP {\bf 0605}, 001 (2006).

\bibitem{JM06c} J. W. Moffat, to be published in {\it Proceedings of the Albert Einstein Century International Conference}, UNESCO, Paris, July 2005; arXiv: astro-ph/0603777.

\bibitem{KK06} T. Kai, H. Kozaki, K. Nakao, {\it et al,} Prog. Theor. Phys. {\bf 117}, 229 (2007).

\bibitem{IW06} A. Ishibashi \& R. M. Wald, Class. Quant. Grav. {\bf 23}, 235 (2006).

\bibitem{AR06} A. E. Romano, Phys. Rev. {\bf D 75}, 043509 (2007).

\bibitem{SR04b} S. R\"as\"anen, JCAP {\bf 0402}, 003 (2004).

\bibitem{BC06} G. Bene, V. Czinner \& M. Vas\'uth, Mod. Phys. Lett. {\bf A 21}, 1117 (2006).

\bibitem{MB05} P. Martineau \& R. Brandenberger, arXiv: astro-ph/0510523 (2005).

\bibitem{DL00} P. S. Drell, T. J. Loredo \& I. Wasserman, ApJ {\bf 530}, 593 (2000).

\bibitem{IN02} H. Iguchi, T. Nakamura \& K. Nakao, Prog. Theor. Phys. {\bf 108}, 809 (2002).

\bibitem{DG06} D. Garfinkle, Class. Quant. Grav. {\bf 23}, 4811 (2006).

\bibitem{SJ01} J. Stelmach \& I. Jakacka, Class. Quant. Grav. {\bf 18}, 2643 (2001).

\bibitem{CR06} D. J. H. Chung \& A. E. Romano, Phys. Rev. {\bf D 74}, 103507 (2006).

\bibitem{EM06} K. Enqvist \& T. Mattsson, JCAP  {\bf 0702}, 019 (2007).

\bibitem{KM05b} E. W. Kolb, S. Matarrese, A. Notari \& A. Riotto, arXiv: hep-th/0503117 (2005). 

\bibitem{GC05} G. Geshnizjani, D. J. H. Chung \& N. Afshordi, Phys. Rev. {\bf D 72}, 023517 (2005).

\bibitem{KM05c} E. W. Kolb, S. Matarrese, A. Notari \& A. Riotto, Mod. Phys. Lett. {\bf A 20}, 2705 (2005).

\bibitem{KM05d} E. W. Kolb, S. Matarrese \& A. Riotto, arXiv: astro-ph/0506534 (2005).

\bibitem{TB05} T. Buchert, Class. Quant. Grav. {\bf 22}, L113 (2005).

\bibitem{SR06b} S. R\"as\"anen, Class. Quant. Grav. {\bf 23}, 1823 (2006).

\bibitem{TB06} T. Buchert, Class. Quant. Grav. {\bf 23}, 817 (2006).

\bibitem{BL06} T. Buchert, J. Larena \& J. M. Alimi, Class. Quant. Grav. {\bf 23}, 6379 (2006).

\bibitem{PS06} A. Paranjape \& T. P. Singh, Class. Quant. Grav. {\bf 23}, 6955 (2006).

\bibitem{RM05} R. Mansouri, arXiv: astro-ph/0512605 (2005), astro-ph/0601699 (2006).

\bibitem{IO04} I. Iwata, K. Ohta, K. Nakanishi, {\it et al}, in {\it Nearby Large-Scale Structures and the Zone of Avoidance} ASP Conference Series, Vol. CS-329, edited by A.P. Fairall and P. A. Woudt (2004).

\bibitem{PM84} M. H. Partovi \& B. Mashhoon, ApJ {\bf 276}, 4 (1984).

\bibitem{HM97} N. P. Humphreys, R. Maartens \& D. R. Matravers, ApJ {\bf 477}, 47 (1997).

\bibitem{AA06} H. Alnes \& M. Amarzguioui, Phys. Rev. {\bf D 74}, 103520 (2006); Phys. Rev. {\bf D 75}, 023506 (2007).

\bibitem{MH97} N. Mustapha, C. Hellaby \& G. F. R. Ellis, MNRAS {\bf 292}, 817 (1997).

\bibitem{LH06} T. H. C. Lu \& C. Hellaby, arXiv: gr-qc/0705.1060 (2007).



\end{thebibliography}
\end{document}